\documentclass[amssymb, showpacs, showkeys, aps, prl, superscriptaddress, twocolumn, longbibliography]{revtex4-2}
\usepackage[colorlinks,bookmarks=false,citecolor=blue,linkcolor=red,urlcolor=blue]{hyperref}

\usepackage[all]{hypcap}
\usepackage{xcolor}
\usepackage{slashed}
\usepackage{graphicx}
\usepackage{amsmath}
\usepackage{latexsym}
\usepackage{epstopdf}
\usepackage{amsmath}
\usepackage{enumitem}
\usepackage{amssymb,amsmath}
\usepackage{multirow}

\usepackage{mathtools}
\usepackage{bbm}
\usepackage{bm}
\usepackage{amsfonts}
\usepackage{amssymb}
\usepackage{calligra}
\usepackage{calrsfs}
\usepackage{makecell}
\usepackage[normalem]{ulem}
\usepackage[USenglish,american]{babel}
\usepackage{physics}
\usepackage{braket}

\newcommand{\be}{\begin{equation}}
\newcommand{\ee}{\end{equation}}
\newcommand{\ba}{\begin{eqnarray}}
\newcommand{\ea}{\end{eqnarray}}
\newcommand{\bs}{\begin{split}}
\newcommand{\es}{\end{split}}

\definecolor{darkblue}{rgb}{0.0,0.0,0.4}
\definecolor{darkgreen}{rgb}{0.0,0.4,0.0}
\definecolor{darkred}{rgb}{0.6,0.0,0.0}

\usepackage{comment}

\begin{document}

\title{Self-bound monolayer crystals of ultracold polar molecules}

\author{Matteo Ciardi}
\email{matteo.ciardi@tuwien.ac.at}
\affiliation{Institute for Theoretical Physics, TU Wien, Wiedner Hauptstraße 8-10/136, 1040 Vienna, Austria}

\author{Kasper Rønning Pedersen}
\email{kasper.pedersen@tuwien.ac.at}
\affiliation{Institute for Theoretical Physics, TU Wien, Wiedner Hauptstraße 8-10/136, 1040 Vienna, Austria}

\author{Tim Langen}
\email{tim.langen@tuwien.ac.at}
\affiliation{Vienna Center for Quantum Science and Technology,
Atominstitut, TU Wien, Stadionallee 2, 1020 Vienna, Austria}

\author{Thomas Pohl}
\email{thomas.pohl@itp.tuwien.ac.at}
\affiliation{Institute for Theoretical Physics, TU Wien, Wiedner Hauptstraße 8-10/136, 1040 Vienna, Austria}

\begin{abstract}
{
We investigate the physics of ultracold dipolar molecules using path-integral quantum Monte Carlo simulations, and construct the complete phase diagram extending from weak to strong interactions and from small to mesoscopic particle numbers. Our calculations predict the formation of self-bound quantum droplets at interaction strengths lower than previously anticipated. For stronger interactions, the droplet continuously loses superfluidity as correlations develop, and is eventually found to undergo a transition to a crystalline monolayer that remains self-bound without external confinement. The spontaneous formation of such two-dimensional phases from a three-dimensional quantum gas is traced back to the peculiar anisotropic form of the dipole-dipole interaction generated by microwave-dressing of rotational molecular states. For sufficiently large particle numbers, crystallization takes place for comparably low interaction strengths that do not promote two-body bound states and should thus be observable in ongoing experiments without limitations from three-body recombination.}
\end{abstract}

\maketitle
Explorations of dipolar quantum matter have revealed a rich phenomenology of exotic quantum states that emerge from the long-range nature and anisotropy of dipole-dipole interactions~\cite{Lahaye2009, Baranov2012}.
Recent advances have been spurred by experiments on ultracold gases of atoms with large magnetic dipole moments \cite{Bottcher2020, Chomaz2022}, which demonstrated the formation of new states such as free-space quantum droplets \cite{Schmitt2016,Chomaz2016} and supersolid phases \cite{Bottcher2019a,Tanzi2019,Chomaz2019}, and probed their physical properties \cite{Petter2019,Hertkorn2021fluctuations,Guo2019,Natale2019,Tanzi2019excitations,Sanchez-Baena2023}. Experiments also highlighted the importance of interaction effects beyond simple mean field theory \cite{kadau2016,Ferrier-Barbut2016} that led to an improved understanding of the role of quantum and thermal fluctuations in weakly interacting quantum gases \cite{Lima2011, Lima2012, Baillie2016,Aybar2019,Sanchez-Baena2023} and their successful modeling in terms of the so-called extended Gross-Pitaevskii equation \cite{Wachtler2016a, Bisset2016, Wachtler2016b, Saito2016}.  

Since electric dipole-dipole interactions generally exceed interactions of atomic magnetic dipoles significantly, ultracold polar molecules have long been anticipated to open new parameter regimes for dipolar quantum matter \cite{Ni2008, Carr2009, Schmidt2022} and for studying quantum magnetism \cite{Yan2013, Li2023}. 
Here, the application of static and microwave fields offers broad opportunities to engineer and control molecular interactions~\cite{Avdeenkov2006,Buchler2007, Gorshkov2008, Karman2018, Lassabliere2018} by properly coupling their rotational states. Such techniques have been designed and used experimentally to stabilize molecular gases against short-range collisional losses \cite{Matsuda2020, Anderegg2021, Li2021, Bigagli2023, Chen2023, Lin2023, Park2023} and ultimately enabled the realization of Bose-Einstein condensates of heteronuclear polar molecules \cite{Bigagli2024}. Recent theory predicts \cite{Schmidt2022, Jin2024, Langen2025}  the formation of molecular droplet states that are similar to those of weakly interacting atomic gases but with different geometries.

\begin{figure}[t!]
  \begin{center}
  \includegraphics[width=\linewidth]{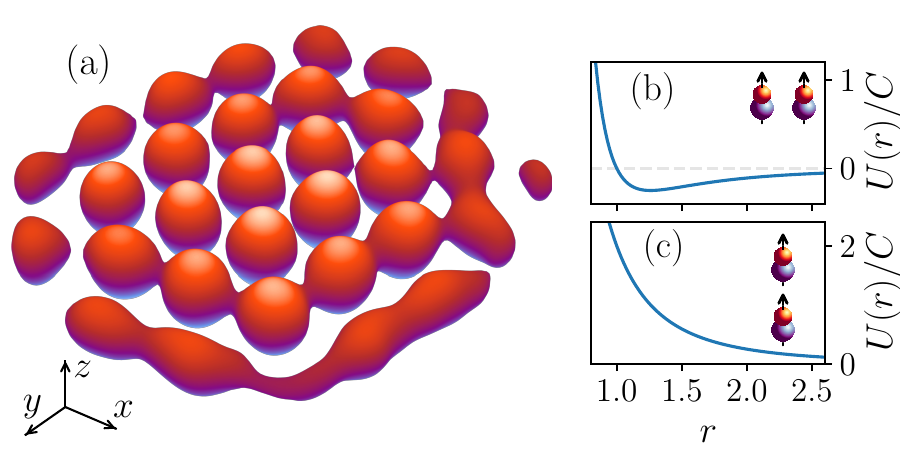}
    \caption{(a) Illustration of a self-bound monolayer crystal, with a single molecule in each lattice site. Shown is the isosurface at a fixed particle density obtained from Quantum Monte Carlo simulations for $N=32$ molecules with a dimensionless interaction strength $C=25$ [cf. Eq.(\ref{eq:potential_reduced})]. Panels (b) and (c) Dipolar potential as a function of the intermolecular distance $r$, in (b) the $xy$ plane and (c) along the $z$ axis.} \label{fig1}
  \end{center}
\end{figure} 

In this letter, we report path integral Monte Carlo (PIMC) simulations of microwave-dressed bosonic molecules and explore the transition from weak to strong dipole-dipole interactions. Our results reveal a characteristic change in the morphology of dipolar quantum droplets towards a two-dimensional liquid, reflecting a transition into the strong-interaction regime. At even stronger interactions, we eventually find a crystallization transition into a self-bound two-dimensional lattice of molecules. Here, particles crystallize into a single monolayer of ordered dipoles, which shares similarities with two-dimensional van der Waals materials \cite{Novoselov2016} and forms in free space without additional confinement due to an in-plane minimum of the interaction potential [Fig.\ref{fig1}(b)] and strong out-of plane repulsion [Fig.\ref{fig1}(c)]. Our calculations indicate that this new phase of dipolar matter should be observable under typical conditions of ongoing experiments with ultracold molecules. Importantly, we find that crystallization occurs for interaction strengths below the threshold for two-body field-linked bound states \cite{supplemental, Avdeenkov2006,Huang2012}, avoiding detrimental three-body losses in experiments \cite{Stevenson2024}. Interestingly, this scenario is akin to the unique behavior of low-temperature helium, where a solid phase emerges directly from an atomic quantum fluid \cite{Dugdale1953}, without the prior formation of a gas or fluid of bound pairs (molecules). Our results thus indicate the great potential of ultracold microwave-dressed molecules for exploring exotic many-body phenomena in quantum fluids with highly controllable interactions, which may enable the experimental realization of long-elusive phases such as defect-induced supersolids \cite{Balibar2010,Cinti2014}.

\begin{figure}[t!]
  \begin{center}
  \includegraphics[width=\linewidth]{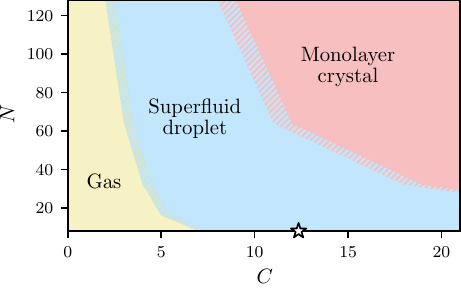}
    \caption{Diagram showing the different phases of molecules in free space for different particle numbers $N$ and interaction strengths $C$. The yellow region indicates the gas phase, the blue region indicates the self-bound, superfluid droplet phase, while the red region indicates the self-assembled mono-layer crystal. Transition regions are represented by striped areas. The star marks the lowest value of $C$ at which the potential admits a bound state (see \cite{supplemental}).}
    \label{fig2}
  \end{center}
\end{figure} 

Among the different approaches to control molecular interactions with external fields, we focus here on the simplest configuration of a single microwave used to couple rotational molecular states and thereby induce a controllable dipole-dipole interaction. Importantly, the dressed molecules also feature a repulsive core that shields pairs of molecules from short-range collisions \cite{Karman2018}. For weak dressing, the resulting interaction potential can be approximated by the simple expression \cite{Deng2023}
\begin{equation} \label{eq:potential}
V(\textbf{R}) = \frac{C_6}{R^6}\sin^2 \theta (1 + \cos^2 \theta) + \frac{C_3}{R^3}(3 \cos^2 \theta -1 )\,,
\end{equation}
where $R = |{\bf R}|$ denotes the distance between two molecules and $\theta$ is the angle between the distance vector ${\bf R}$ and the orientation of the dipole, set by the applied microwave field. The interaction coefficients $C_3$ and $C_6$ are determined by the molecular dipole moment and can be controlled by the parameters of the applied microwave field~\cite{supplemental}.

Choosing the dipole orientation along the $z$-axis, the interaction potential is purely repulsive along this direction [Fig.\ref{fig1}(b)] but features a minimum in the $x-y$ plane [Fig.\ref{fig1}(c)] that is set by the competition of the short-range repulsion and the long-range, attractive dipole-dipole interaction. The associated distance $R_0 = (C_6 / C_3)^{1/3}$, at which the potential $V(R_0)=0$ vanishes, provides a natural unit of length. Upon scaling energies by $E_0 = \hbar^2 / (m R_0^2) $, and introducing $\textbf{r} = \textbf{R} / R_0$, the Hamiltonian for $N$ interacting molecules with mass $m$ can be written as
\begin{equation} \label{eq:hamiltonian_reduced}
\hat{H} = - \frac{1}{2} \sum_{i=1}^{N} \nabla_{{\bf r}_i}^2 + \sum_{i<j} U({\bf r}_i-{\bf r}_j),
\end{equation}
with the dimensionless interaction potential 
\begin{equation} \label{eq:potential_reduced}
U({\bf r}) = \frac{V({\bf R})}{E_0}= C \left( \frac{\sin^2 \theta (1 + \cos^2 \theta)}{r^6} + \frac{3 \cos^2 \theta -1}{r^3} \right),
\end{equation}
that depends on a single interaction parameter $C = m C_3^{4/3} / (\hbar^2 C_6^{1/3})$. Hence, the equilibrium properties of the system are determined by only two parameters, the particle number $N$ and the dimensionless interaction strength $C$.

We study the low-temperature quantum phases of the Hamiltonian \eqref{eq:hamiltonian_reduced} by using path integral Monte Carlo simulations (PIMC) \cite{ceperley1995}. Quantum Monte Carlo methods have been successfully applied to the description of dipolar bosonic systems in different geometries \cite{Nho2005, Mora2007, Jain2011, Macia2011, Cinti2017a, Cinti2017b, Saito2016, Macia2016, Bottcher2019b, Boninsegni2021, Bombin2024a, Ghosh2024, Langen2025, Zhang2025}. In PIMC, the partition function is sampled as a Monte-Carlo integral over a set of discrete imaginary-time trajectories, i.e. world lines associated to each particle. An efficient representation of the symmetrized many-body quantum state is possible through the so-called worm algorithm \cite{Boninsegni2006} that samples bosonic permutations by opening, closing and permuting world lines.  

Just like other Quantum Monte Carlo methods, PIMC simulations should, in principle, yield exact results, apart from statistical uncertainties. In practice, however, a judicious choice of initial configurations and a proper balance of different Monte Carlo moves is often required to achieve efficient convergence to equilibrium. As we discuss below, this is particularly important in the present case for a reliable prediction of phase boundaries and to avoid unphysical states associated with metastable world line configurations. Our simulations are performed at a low temperature $T = 0.01E_0/k_B$, and we have confirmed that further decreasing the temperature does not change the results reported here. Our results across a range of values of $C$ and $N$ are summarized in Fig.~\ref{fig2}.

We simulate a finite number of particles in free space with open boundary conditions. Hence, the equilibrium phase for weak interactions $C\ll1$ corresponds to an entirely delocalized gas with vanishing density and a finite kinetic energy proportional to the temperature $T$. As the value of $C$ increases, the attractive part of the dipole-dipole interaction potential counteracts the kinetic energy of the particles and eventually promotes the formation of self-bound quantum droplets, as previously found in Bose-Einstein condensates of magnetic atoms \cite{Schmitt2016, Chomaz2016} and predicted for molecular interactions such as Eq.~(\ref{eq:potential}).

\begin{figure}[b!]
  \begin{center}
  \includegraphics[width=\linewidth]{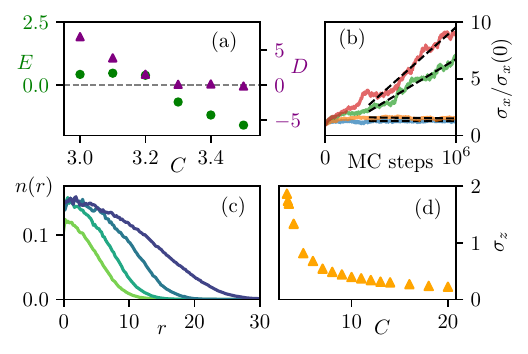}
    \caption{(a) Total energy, $E$,  (green circles) and slope, $D$, of $\sigma_x$(purple triangles) for $N=64$ molecules at different values of $C$ around the gas-to-droplet transition. The latter is obtained by fitting the size $\sigma_x$ as a function of  Monte Carlo steps, as illustrated in panel (b) for $C=2.5$ (red), 3.0 (green), 3.5 (orange), 4.0 (blue). (c) Transverse droplet size $\sigma_z$ as a function of $C$ at $N=64$. (d) Radial two-dimensional density $n=\int{\rm d}z\:n_{\rm 3D}({\bf r})$ of the droplet for $N=16$ (green), $N=32$ (light blue), $N=64$ (dark blue), $N=128$ (purple) at $C$ close to the gas transition.}
    \label{fig3}
  \end{center}
\end{figure}

While the Monte Carlo simulations straightforwardly yield stable droplet states for sufficiently large interactions, an accurate determination of the boundary between the gas and droplet phase requires further consideration. To this end, we start from a localized initial configuration and track the changing state of the system during Monte Carlo moves by monitoring the spatial spread of the molecular gas, defined as $ \sigma_x^2 ={N}^{-1} \int d\textbf{r} \, x^2 n_{\rm 3D}(\textbf{r})$, and analogously for $\sigma_y$ and $\sigma_z$, where $n_{\rm 3D}$ is the density of molecules. In the droplet phase, the spatial spread converges to a finite value following an initial relaxation phase. In the gas phase, the initial configuration expands continuously, and $\sigma_x$ asymptotically grows linearly with the number of Monte Carlo steps. The slope, $D$, of this linear growth can be used to discriminate between the gas phase ($D>0$) and the droplet phase ($D=0$), as illustrated in Fig.~\ref{fig3}(a); the transition point is confirmed by the energy turning negative in the droplet phase. Plots of the size as a function of Monte Carlo steps are shown Fig.~\ref{fig3}(b).

In order to facilitate convergence of the simulations towards the droplet phase, we sample the starting position of the particles from a uniform distribution inside a cylinder of a given radius and height. We have thoroughly checked that the results do not depend on the choice of the initial cylinder \cite{supplemental}. However, we observe that if the initial configuration is too dilute, the molecules tend to fragment into smaller separate droplets. While such multi-droplet states may appear to form an intriguing metastable phase \cite{Langen2025}, we find that they disappear upon including collective Monte Carlo moves of multiple particles \cite{supplemental}. Similar behavior was also found in the presence of harmonic confinement, indicating that long-lived multi-droplet states do not represent the equilibrium state under the interaction Eq.(\ref{eq:potential}) \cite{supplemental}, although multidroplet states may form in experiments in out-of-equilibrium conditions. 

In order to ensure the accuracy of our phase boundary for the gas-to-droplet transition, we also run simulations using as starting configuration the most weakly bound droplets observed at each $N$, while reducing $C$. In each case, we observe the disgregation of the droplet, confirming the result from the cylinder starting conditions. The critical interaction strengths are notably smaller than recent predictions, indicating that the described approach succeeds in finding lower energy states than previous variational estimates \cite{Jin2024} and Monte Carlo simulations \cite{Langen2025}.

\begin{figure*}[t]
  \begin{center}
  \includegraphics[width=\linewidth]{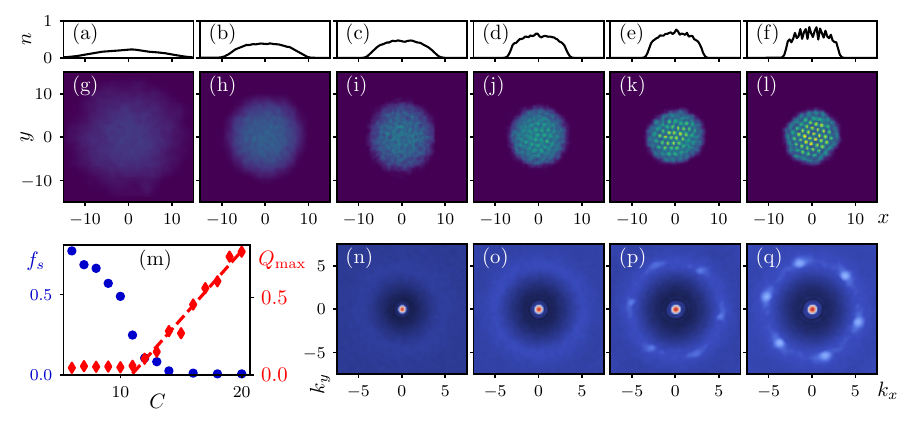}
    \caption{Structural properties of the droplet in the strongly interacting regime. (a--f) Profiles of the 2D density $n$ for $N=64$ molecules at successive values of $C= 4$, $6$, $8$, $12$, $14$, and $18$. (g--l) $n$ as a function of both $x$ and $y$, for the same values of $C$. (m) Contrast $Q_{\rm max}$ (red diamonds) and superfluid fraction (blue circles) as a function of $C$. (n--q) Structure factor $S$ for $C= 8$, $12$, $14$, and $18$, respectively.}
    \label{fig4}
  \end{center}
\end{figure*}

Close to the transition, the droplets feature a superfluid fraction close to unity and a quasi-2D geometry, extended in the $x-y$ plane. The pancake-like shape arises due to the anisotropic form of the interaction potential\eqref{eq:potential_reduced}, where the long-range attraction perpendicular to the dipole-orientation [Fig.\ref{fig1}(b)] facilitates self-trapping in the $x-y$ plane, while the strong longitudinal repulsion [Fig.\ref{fig1}(c)] confines the droplet along the $z$-direction. The central density remains constant along the transition line [Fig.\ref{fig3}(c)], but increases away from it upon increasing the interaction strength $C$ [Fig\ref{fig4}(a--f)]. Near the transition, the droplet size along the $z$-axis, $\sigma_z>1$, exceeds the characteristic scales of the interaction potential and the droplet behaves as a quasi-2D fluid with an approximately Gaussian density profile along all directions [Fig.\ref{fig3}(c) and (d)]. However, $\sigma_z$ decreases at stronger interactions, giving rise to the formation of an effectively two-dimensional droplet with a near-constant bulk density and a decreasing superfluid fraction with rising $C$ [Fig.\ref{fig4}(m)]. This distinctively different geometry and lower superfluidity indicate a new droplet phase of strongly correlated molecules.

Further increasing the interaction strength continually increases the droplet density, and the superfluidity eventually vanishes at a 2D droplet density of $n \approx 0.7$ [Fig.\ref{fig4}(m)], for which the interparticle spacing corresponds to the minimum of the in-plane potential well at $r_{\rm min}=2^{1/3}$ seen in Fig.\ref{fig1}(b). At this point, the system undergoes a transition to an insulating crystalline monolayer of molecules, as illustrated by the density profiles shown in Fig.\ref{fig4}(g--l). In order to quantify the crystallization transition, we define the structure factor
\begin{equation}
    S(\textbf{k}) = \left| \int d\textbf{r} \, \rho(\textbf{r}) e^{- i \textbf{k} \cdot \textbf{r}} \right|^2 \,, 
\end{equation}
where we only consider in-plane momenta with $k_z=0$. Upon entering the crystalline monolayer phase, $S({\bf k})$ develops a peak structure at the lattice momentum $|{\bf k}|\sim2 \pi / r_{\rm min} \approx 5$, as shown in Fig.\ref{fig4}(n--q). Its six-fold rotational symmetry reflects the triangular lattice of single molecules that forms in the bulk of the 2D droplet. We calculate the average $\bar{S}(k)=\frac{1}{2\pi}\int {\rm d}\varphi S(\textbf{k})$ to determine the angular contrast $Q(k)=\frac{1}{2\pi} \int {\rm d}\varphi ( S({\bf k}) - \bar{S}(k))^2$, by integrating over the angle $\varphi$ in the $k_x-k_y$ plane. One can then use the maximum contrast $Q_{\rm max}={\rm max}_k Q(k)$ to characterize the crystalline order of the droplet. As shown in Fig.\ref{fig4}(m), the order parameter $Q_{\rm max}$ starts to increase linearly with $C$ coincidentally with the vanishing of the superfluidity. We can thus use a linear fit to determine the transition point, as illustrated in the figure. 

The transition line between the superfluid droplet and the monolayer-crystal phase is shown in Fig.\ref{fig2}. The critical value of $C$ decreases with the particle number and is expected to approach a constant value as $N$ increases towards the thermodynamic limit. For sufficiently large particle numbers $N\gtrsim 60$, the crystallization occurs at interaction strengths $C<C_b\sim12.3$ for which the potential does not promote two-body bound states. Remarkably, the considered system of microwave-dressed molecules thus resembles the rare phenomenology of low-temperature helium, where a solid phase occurs via crystallization of an atomic superfluid without the formation of a molecular gas or fluid of two-body bound states. Interestingly, the two-body bound state is far more extended along the $z$-direction \cite{supplemental} than the spatial spread $\sigma_z$ of the droplet. The presence of the bound state is thus not expected to impact the physics of the droplet phase for the parameters considered in Fig.\ref{fig2}.

The absence of bound states is also important for potential experiments, as it should enable the observation of the predicted monolayer-crystal phase without detrimental losses from three-body recombination \cite{Stevenson2024,Chen2024} into field-linked tetramer states of two bound molecules. Microwave dressing and shielding have been demonstrated in recent experiments with ultracold CaF \cite{Anderegg2021}, NaK \cite{Schindewolf2022}, NaRb~\cite{Lin2023} and NaCs \cite{Bigagli2023, Bigagli2024} molecules, using different field configurations. 
The simplest scheme based on a single field yields an interaction potential that is approximately given by Eq.(\ref{eq:potential_reduced}), where the dimensionless interaction strength, $C$, is determined by the molecular dipole moment and mass, as well as the frequency detuning and Rabi frequency of the applied microwave \cite{supplemental}.  Typical microwave Rabi frequencies of $10$MHz \cite{Anderegg2021,Bigagli2023,Lin2023,Schindewolf2022} yield interaction strengths of up to $C\sim11$ for NaK molecules and even stronger interactions with up to $C\sim55$ for NaCs molecules, such that the predicted phases appear to be within experimental reach for both molecular species. 

In summary, we have explored the phases of bosonic ultracold dipolar molecules in free space. The peculiar form of the interaction potential that is induced by microwave dressing with a single applied field promotes the formation of self-bound quantum droplets with a characteristic pancake-like geometry. Our PIMC simulations predict this droplet-formation to occur at weaker interactions than previously calculated \cite{Jin2024,Langen2025}. As the droplet solution features a negative total energy, this result is conceptually consistent with the variational calculations of \cite{Jin2024}. The simulations reveal a change in the droplet morphology upon approaching the strongly interacting regime, where the system forms a two-dimensional self-bound fluid that ultimately crystallizes into a monolayer of individual molecules with a triangular lattice structure. Their physical origin and properties thus differ from Wigner crystals of repulsive dipoles under strong three-dimensional confinement \cite{Buchler2007, Astrakharchik2007, Jain2011}.

A careful variation and analysis of the starting conditions, the parameters of the PIMC algorithm, and the introduction of collective Monte-Carlo moves leads us to exclude the formation of multi-droplet states in free space, and we have also found no evidence for such ground-state solutions under external confinement upon a proper choice of Monte-Carlo parameters. Such multi-droplet states or even cluster solids and supersolids may, however, form under different forms of interactions that can be engineered with multiple dressing fields \cite{Bigagli2024,Karman2025} and motivate future studies. 

Our analysis suggests that the predicted monolayer crystals should be observable under typical conditions of ongoing ultracold molecule experiments. In particular, we show that they form for interaction strengths that do not promote two-body bound states and, thus, avoid detrimental effects of collisional losses in potential experiments. This appears reminiscent of the behavior of helium to crystallize directly from a superfluid state without an intermediate molecular gas or liquid phase of two-body bound states. This demonstrated analogy together with the high tunability of interactions indeed presents an exciting outlook for future explorations of exotic and long-elusive phases such as defect-induced supersolids in new interaction regimes that are now becoming accessible with ultracold dipolar molecules. 

\begin{acknowledgments}
We thank Juan S\'{a}nchez-Baena, Jordi Boronat, Fabio Cinti, Tao Shi, and Andreas Schindewolf for helpful discussions.
This work was supported by funding from the
Austrian Science Fund (Grant No. 10.55776/COE1) and
the European Union (NextGenerationEU), by the SNSF
through the Swiss Quantum Initiative, and from the
European Research Council through the ERC Synergy
Grant "SuperWave" (Grant No. 101071882) and the ERC Starting Grant "NEWMAT" (Grant No. 949431).
\end{acknowledgments}

\bibliography{biblio}

\newpage
\clearpage

\setcounter{page}{1}

\section{Supplemental Material}

\subsection{Units and experimental parameters}

As shown in \cite{Deng2023}, the interaction potential between two microwave-dressed molecules can be approximately calculated from Eq.(1) of the main text, with the interaction coefficients 
\begin{equation}
    C_3 = \frac{d^2\Omega^2}{48 \pi \epsilon_0 (\Omega^2 + \Delta^2)} \,,
\end{equation}
and
\begin{equation}
C_6 = \frac{d^4\Omega^2}{128 \pi^2 \epsilon_0^2 \hbar  (\Omega^2 + \Delta^2)^{3/2}} \,.
\end{equation}
where $d$ is the bare dipole moment of the molecule, while $\Delta$ and $\Omega$ denote the frequency detuning and Rabi frequency of the applied microwave field. With these expressions, one can determine the length scale 
\begin{equation}
    R_0 = \left( \frac{C_6}{C_3} \right)^{\frac{1}{3}} = \left( \frac{3 d^2}{8 \pi \epsilon_0 \hbar \sqrt{\Omega^2 + \Delta^2}} \right)^{\frac{1}{3}}
\end{equation}
and the dimensionless interaction strength
\begin{equation} \label{eq:apx_C_long}
    C = \frac{C_3 m}{\hbar^2  R_0} =\frac{ m\Omega^2}{48 \pi \epsilon_0 \hbar^2} \left( \frac{8 \pi \epsilon_0 \hbar d^4 }{3 (\Omega^2 + \Delta^2)^{5/2}} \right)^{\frac{1}{3}}
\end{equation}
defined in the main text. 
For NaK molecules with a dipole moment $d \approx 2.85$ D and a mass $m \approx 62$ amu, one obtains $C\approx11$ and $R_0\approx57$ nm for resonant ($\Delta=0$) driving with a Rabi frequency $\Omega/2\pi=10$ MHz, which is typical for experiments \cite{Anderegg2021,Bigagli2023,Lin2023,Schindewolf2022}.  For a NaCs molecule, which has a larger dipole moment $d \approx 4.7$ D and a larger mass $m \approx 156$ amu, the same microwave field yields a larger interaction strength $C \approx 55$ and length scale $R_0=80$ nm. One can reduce the interaction strength by increasing the detuning or lowering the Rabi frequency.  For example, for off-resonant driving ($\Delta=1.5\Omega$) with  $\Omega/2\pi=1.5$ MHz, one generates an interaction strength $C\approx11$ and a somewhat shorter unit length $R_0\approx125$ nm, which is in the range of typical interparticle spacings in cold atom experiments. 

\subsection{Independence from initial configuration}
\begin{figure}[t!]
  \begin{center}
  \includegraphics[width=\linewidth]{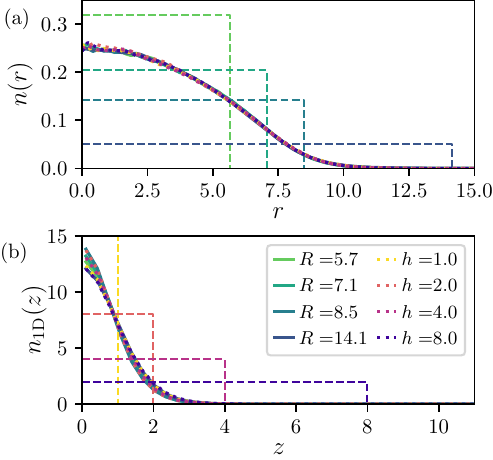}
    \caption{Final density profiles, for simulations of $N=32$ particles at $C=5.5$ with different starting conditions (various values of $R$ at $h=0$, and various values of $h$ at $R=8.5$). (a): Radial density profile in the $xy$ plane. (b): Density profile integrated along the $x$ and $y$ axes. The rectangles in both plots represent the average densities associated with the respective starting configurations.}
    \label{fig5}
  \end{center}
\end{figure}

In our simulations, we choose the starting configuration by picking $N$ particle positions at random from a distribution $\rho(\textbf{r})$ which is uniform inside a cylinder of radius $R$ and height $h$, and null outside of it. To be reliable, the results of PIMC simulations must not depend on the starting configuration chosen. To show that this is in fact the case, we display some examples in Fig.\ref{fig5}. In this case, we have performed eight simulations with starting configurations sampled from different disks: four disks with increasing radius at $h=0$, and four more with increasing heights at fixed $R$. In all cases, we plot the density profiles obtained after the various simulations have converged. The results are summarized in Fig.\ref{fig5} for different combinations of $R$ and $h$, and demonstrate that the simulations converge to a common droplet state. 

While performing the simulations at different starting densities, we found that the independence on starting conditions breaks down when the starting density becomes too low. In this case, particles may come together to form local agglomerates, but do not proceed to form a single droplet. While this could be attributed to being in a different phase due to the lower density, the total energy measured in this case is significantly higher compared to that of the droplet displayed in Fig.\ref{fig5}. This indicates a metastable configuration, which should eventually converge to the correct equilibrium state, i.e. a single droplet. However, rapid convergence can often be hindered due to general limitations of the QMC approach. This issue is solved by the introduction of cluster moves, as briefly described below.

\begin{figure*}[t!]
  \begin{center}
  \includegraphics[width=\linewidth]{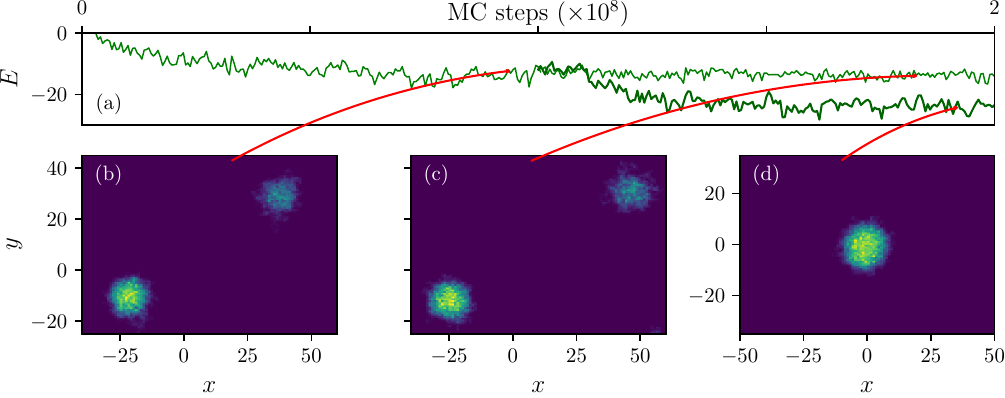}
    \caption{Density profiles in the center of mass frame of reference, for a simulation of $N=64$ particles with starting density $5 \times 10^{-3}$ ($R=64$). (a): Density after $10^8$ Monte Carlo steps with the standard algorithm. (b): Density after evolving (a) for another $10^8$ Monte Carlo steps with the standard algorithm. (c): Density after evolving (a) for another $10^8$ Monte Carlo steps, including cluster moves.}
    \label{fig6}
  \end{center}
\end{figure*} 

\subsection{Monte Carlo moves}
While performing various kinds of simulations, we encountered different problems related to insufficient sampling by the standard PIMC algorithm. These issues lead the world line configurations to become stuck in metastable configurations of scattered droplets, or connected multi-droplet configurations. Here we detail the most prominent issues encountered and how to solve them, leading in all cases to establishing a single droplet as the equilibrium state with a lower total energy. We note that such states are distinct from potential metastable states that may appear in experiments, which have a dynamical origin and have been observed in experiments with atomic dipolar gases~\cite{Schmitt2016}.

\begin{figure}[b!]
  \begin{center}
  \includegraphics[width=\linewidth]{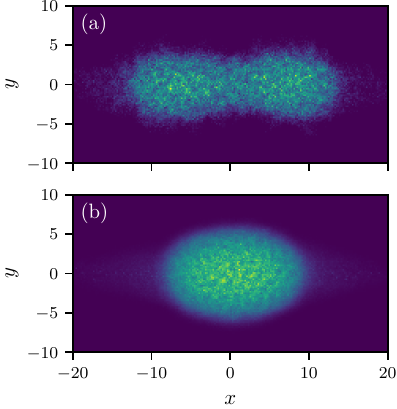}
    \caption{Simulations at $C=9$, $l=2$. (a): Density profile for a simulation with $W=1$, after $10^8$ Monte Carlo steps. Continuing the simulation causes only minimal changes in the configuration. (b): Density profile after the same number of steps, with $W=100$.}
    \label{fig7}
  \end{center}
\end{figure}
\subsubsection{Absence of multi-droplets configurations at equilibrium}
If the initial density is too low, the molecules will separate into two or more fully separate droplets, which appear to be balanced in a metastable configuration. This occurs because traditionally PIMC algorithms employ local moves, changing one particle at a time; when interactions are too strong, it becomes extremely improbable for these local moves to displace a single world line from one droplet to the other, as the potential barrier to remove an entire particle is too large. Introducing droplet-wide moves, which modify or displace all world lines in a droplet at once, allows the droplets to move towards each other and eventually combine. An example of this phenomenon is shown in Fig.\ref{fig6}. The starting density profile in this case is a disk with height $h=0$ and radius $R=64$, extending beyond the limits of the plot. The top panel represents the total energy estimator as a function of Monte Carlo steps, which, with only local moves, converges to a certain value around $E \approx 17$. The introduction of multi-particle moves in this case has a stark effect, allowing the fragmented clusters to combine into a single droplet with lower energy, the dynamics represented by the thicker line. As shown in the lower panels, the density distribution is essentially fixed after a certain point when using only local moves in a two-droplet configuration. Introducing global moves allows the two droplets to come closer and eventually merge.

\begin{figure}[b!]
  \begin{center}
  \includegraphics[width=\linewidth]{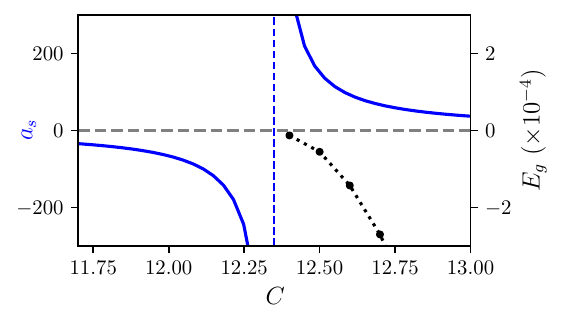}
    \caption{Scattering length (in units of $R_0$) as a function of the interaction parameter $C$ (blue solid line) shown in the vicinity of the divergence at $C_b \approx12.35$. The dotted black line shows the ground state energy $E_g$ calculated for $C>C_b$.}
    \label{fig8}
  \end{center}
\end{figure}

\subsubsection{Efficiency of worm moves}
In PIMC simulations using the worm algorithm, configurations with closed and open worldlines are both sampled. Closed worldlines contribute to diagonal observables such as density, superfluid density, and structure factor, while open worldlines contribute to off-diagonal observables like the one-body density matrix. The simulation includes a free parameter ($C$ in \cite{Boninsegni2006}, which we call $W$ here to avoid confusion with the interaction), which regulates how much time is spent sampling open worldline configurations. We find that, if $W$ is too small so that the probability to perform worm moves is not sufficiently large, the world lines can become stuck in a single droplet with two peaks and only relax very slowly to the single-droplet ground state. An example, obtained in the presence of harmonic confinement along $y$ and $z$ with oscillator length $l$, is shown in Fig.\ref{fig7}. 

\begin{figure}[t!]
  \begin{center}
  \includegraphics[width=\linewidth]{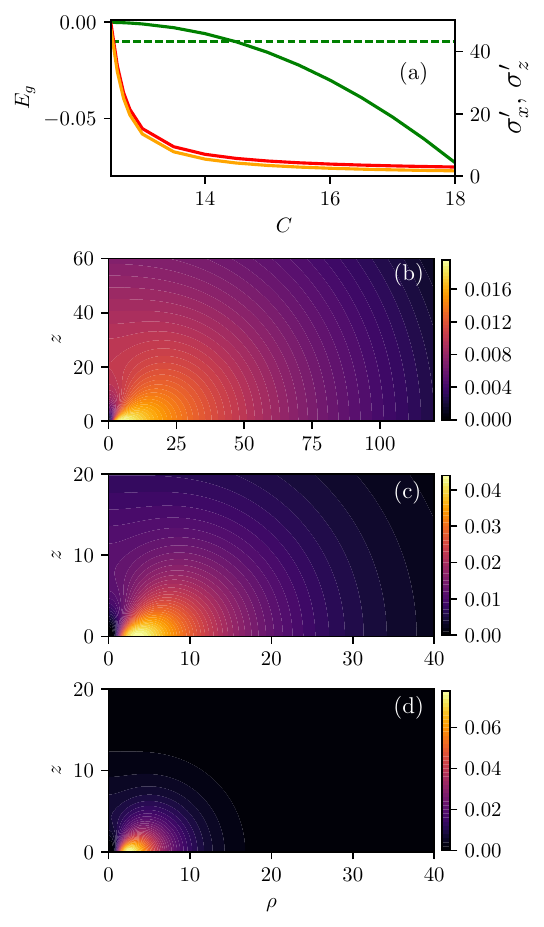}
    \caption{(a): Energy of the bound state (green line) and root mean square sizes of the wavefunction $\sigma'_x$ (red line), $\sigma'_z$ (orange line), as a function of $C$. The sizes decrease but keep being greater than one, also away from the bound state resonance. The dashed line at $-T$ is a visual indicator to compare the bound state energy and the temperature at which the Monte Carlo simulations have been performed. (b)-(d): Probability densities of the wavefunction for $C=12.5, 14, 18$ as a function of the cylindrical coordinates $\rho$ and $z$. The color scale is the same in all three cases.}
    \label{fig9}
  \end{center}
\end{figure}

\subsection{Two-body physics} 
In addition to the many-body Hamiltonian (\ref{eq:hamiltonian_reduced}), we consider scattering between two molecules interacting via the potential (\ref{eq:potential_reduced}). By expanding the wavefunction into partial waves, we diagonalize the Hamiltonian of the relative motion 
\begin{equation} \label{eq:two_body_hamiltonian_reduced}
\hat{H}_r = -\nabla_{{\bf r}}^2 + U({\bf r}),
\end{equation}
and extract the scattering length from the low-energy solution~\cite{Johnson1973}. The Hamiltonian (\ref{eq:two_body_hamiltonian_reduced}) is rescaled by $E_0$ introduced in the main text. Higher partial waves contribute to the scattering length due to the anisotropy of the potential \cite{Yi2001}, which couples every second and every fourth partial wave. As shown in Fig.\ref{fig8}, we find a divergence of the scattering length for $C_b \approx 12.35$, consistent with other studies \cite{Lassabliere2018,Langen2025, Jin2024}, indicating the formation of a field-linked bound state. In order to characterize the resulting bound state, we diagonalize the Hamiltonian (\ref{eq:two_body_hamiltonian_reduced}) for $C > C_b$. The obtained energies are shown in Fig.\ref{fig9}, together with the root mean square sizes $\sigma_x', \sigma_z'$ of the bound-state wavefunction given by 
\begin{equation}
\sigma_a'^2 = \int d\textbf{r} \, \abs{\psi(\textbf{r})}^2 a^2,
\end{equation}
for $a=x,z$. Furthermore, we indicate the temperature at which the Monte Carlo simulations have been performed. As expected, we see a large drop in the size of the wavefunction away from the critical value $C_b$. For $C=18$, well above the resonance, $\sigma_z'$ remains significantly greater than $1$, i.e. large compared to the relevant length scale of the self-bound droplet state. 
\end{document}